# Broadband Source-Surrounded Cloak for On-Chip Antenna Radiation Pattern Protection

Hanchuan Chen, Fei Sun, Yichao Liu, Shuai Zhang, *Fellow, IEEE*, Hongming Fei, and Zhihui Chen

***Abstract*—As the frequency range of electromagnetic wave communication continues to expand and the integration of integrated circuits increases, electromagnetic waves emitted by on-chip antennas are prone to scattering from electronic components, which limits further improvements in integration and the protection of radiation patterns. Cloaks can be used to reduce electromagnetic scattering; however, they cannot achieve both broadband and omnidirectional effectiveness simultaneously. Moreover, their operating modes are typically designed for scenarios where the source is located outside the cloak, making it difficult to address this problem. In this work, we propose a dispersionless air-impedance-matched metamaterial over the 2-8 GHz bandwidth that achieves an adjustable effective refractive index ranging from 1.1 to 1.5, with transmittance maintained above 93%. Based on this metamaterial, we introduce a broadband source-surrounded cloak that can guide electromagnetic waves from a broadband source surrounded by the cloak in any propagation direction to bypass obstacles and reproduce the original wavefronts outside the cloak. Thereby protecting the radiation pattern from distortion due to scattering caused by obstacles. Our work demonstrates significant potential for enhancing the integration density of integrated circuits and improving the operational stability of communication systems.**

## I. INTRODUCTION

Improving electromagnetic wave emission efficiency has become a critical technological objective with diverse applications. In particular, on-chip communication systems benefit from it, resulting in enhanced signal quality and increased transmission reliability[1], [2], [3]. However, as the component integration of on-chip communication systems continues to increase and the communication frequency bands expand, the electromagnetic waves emitted by on-chip antennas are prone to scattering by electronic components/metallic electronic components[4], [5], [6]. This scattering will cause distortion in the radiation pattern, ultimately leading to reduced electromagnetic wave emission efficiency. To balance electromagnetic wave emission efficiency, further improvement in the integration of on-chip communication systems has been limited. Traditional cloaks can be applied around electronic components to reduce their scattering of incident waves from the outside[7], [8], [9]. In recent years, an increasing number of invisibility mechanisms and methods have been proposed. Notable research works in this field include cloaks based on transformation optics[10], [11], [12], [13], cloaks based on optical conformal transformation[14], [15], [16], and cloaks based on scattering-cancellation[17], [18], [19], [20] *et al*. However, current cloaks are affected by two critical problems that limit their application in protecting the radiation patterns of on-chip sources within highly integrated on-chip communication systems. First, there is a significant bandwidth limitation. Most full-space cloaks operate only at single frequencies or narrow frequency bands[21], [22], [23]. Although directionally restricted cloaks[24], [25] or carpet cloaks may achieve broadband operation[26], [27], [28], they are not suitable for protecting the radiation patterns of on-chip sources or antennas. Second, all traditional cloaks are designed for electromagnetic waves incident from the outside to encounter concealed regions, but they do not address scenarios where electromagnetic waves generated by internal sources propagate outward and encounter concealed regions containing strong electromagnetic scatterers. Therefore, there is still a lack of a cloak that allows electromagnetic waves emitted in any propagation direction to bypass obstacles without scattering and operates across a broadband frequency range, thus protecting the radiation patterns of on-chip sources/antennas from distortion caused by surrounding electronic components.

In this work, a dispersionless air-impedance-matched metamaterial (AIMM) operating over the 2-8 GHz bandwidth is proposed, achieving an adjustable effective refractive index ranging from 1.1 to 1.5. This adjustability is achieved by adjusting the dielectric mixing ratio $m$, while maintaining a transmittance $t$ above 93% across the entire bandwidth. We further propose a broadband source-surrounded cloak (SSC) based on AIMM for on-chip radiation pattern protection. The SSC operates with a bandwidth of 6 GHz centered at $f_0$ = 5 GHz ($\lambda_0$ = 60 mm). As illustrated in Fig. 1, the SSC fundamentally differs from traditional cloaks by surrounding the source rather than the scatterer. Electronic components can

This work was supported by the National Natural Science Foundation of China (Nos. 12274317, 12374277), and 2024 Postgraduate Research and Innovation Project of Shanxi Province (2024KY220). *(Corresponding author: Fei Sun.)*

Hanchuan Chen, Fei Sun, Yichao Liu, Hongming Fei and Zhihui Chen are with the Key Lab of Advanced Transducers and Intelligent Control System, Ministry of Education and Shanxi Province, College of Physics and Optoelectronics Engineering, Taiyuan University of Technology, Taiyuan 030024, China (e-mail: sunfei@tyut.edu.cn).

Shuai Zhang is with the Department of Electronic Systems, Aalborg University, 9220 Aalborg, Denmark (e-mail: sz@es.aau.dk).

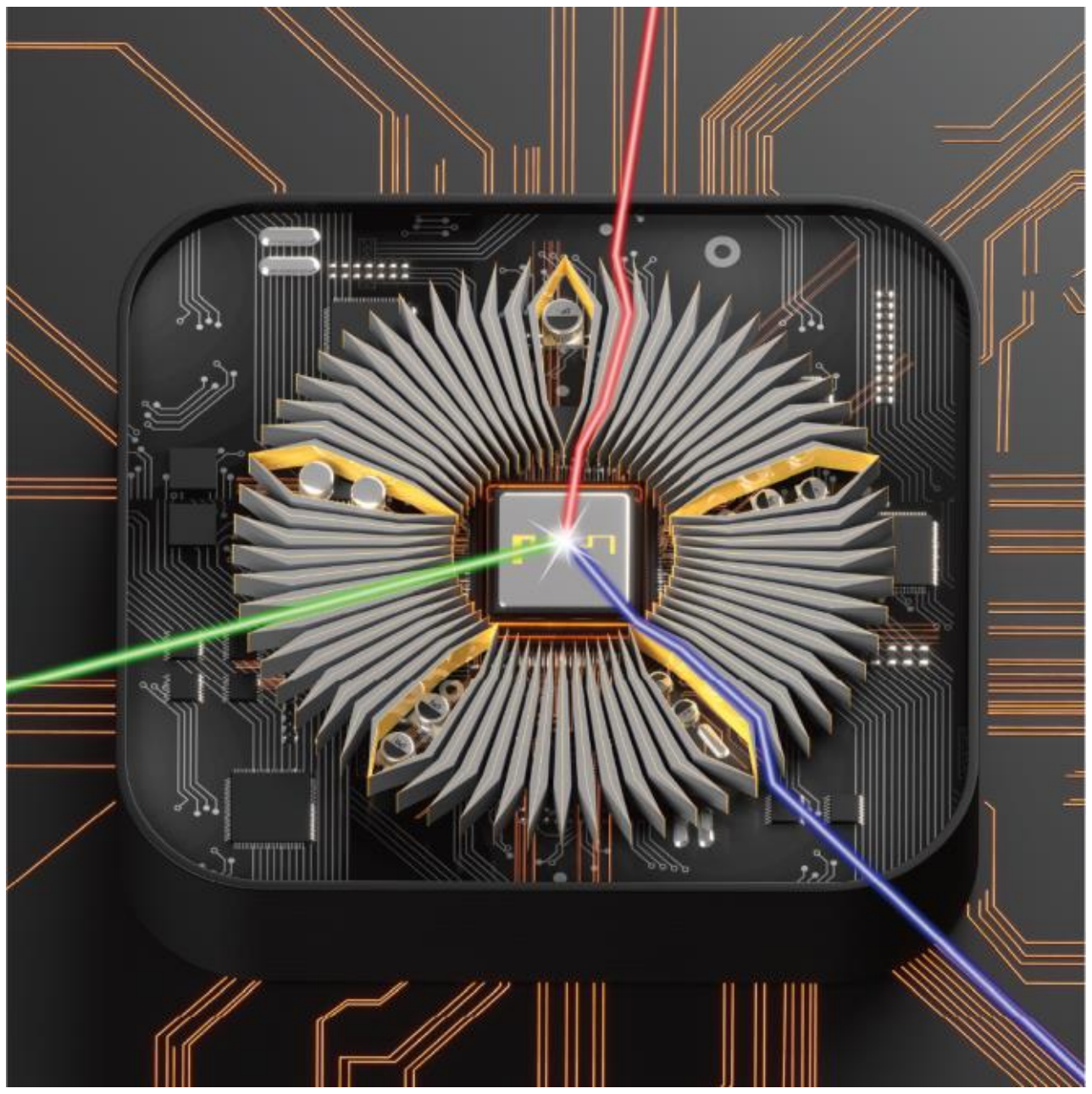

**Fig. 1.** The schematic view of SSC applied to on-chip communication system. It demonstrates the on-chip antenna surrounded by SSC, with obstacles like electronic components that can scatter electromagnetic waves placed in the concealed regions of SSC. The electromagnetic waves emitted by the on-chip antenna in any propagation direction are guided around the obstacles by SSC and exit without distortion.

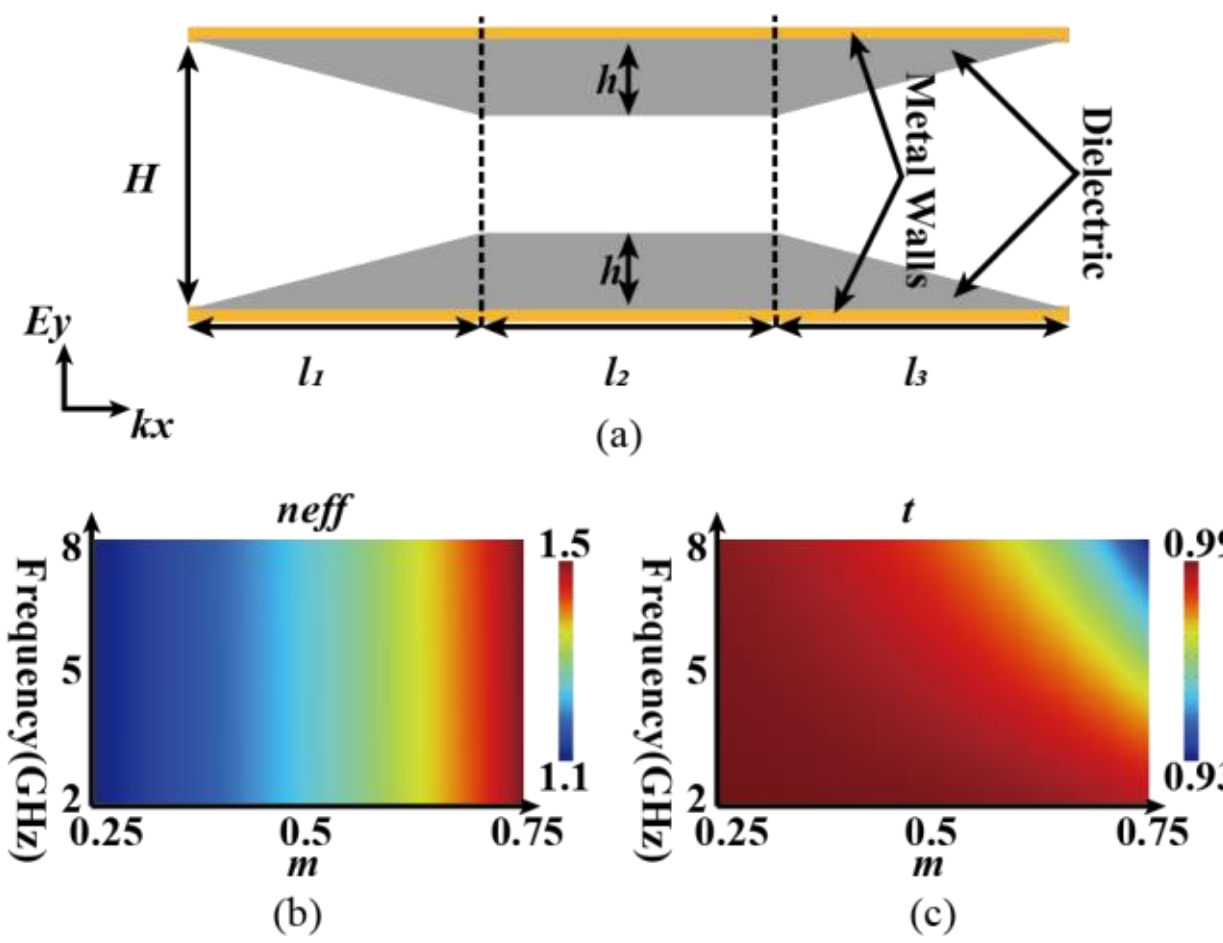

**Fig. 2.** The design of AIMM. (a) The structural schematic diagram of AIMM. The yellow lines represent metal walls, and the gray areas are dielectric regions. (b) Effective refractive index $n_{eff}$ as a function of $m = 2h/H$ and the frequency. (c) Transmittance $t$ as a function of $m = 2h/H$ and the frequency. In (b) and (c), $l_1 = 0.02\lambda_0$, $l_2 = 0.02\lambda_0$, $l_3 = 0.02\lambda_0$, $H = 0.02\lambda_0$, and the relative permittivity of dielectric regions $\varepsilon_{rd} = 10$.

be placed within the concealed regions, while broadband electromagnetic waves emitted omnidirectionally from the source are guided by the SSC to bypass the components without scattering. Crucially, the electromagnetic waves fully reproduce their original wavefronts after exiting the cloak.

## II. Methods And Materials

Fig. 2(a) details the structure of the AIMM, comprising top and bottom metal walls (treated as perfect electric conductor boundaries at microwave frequencies) sandwiching two symmetrical dielectric regions with relative permittivity $\varepsilon_{rd}$. The AIMM operates effectively solely for transverse magnetic (TM) waves, and accordingly, this work is confined to TM-polarized electromagnetic phenomena. The AIMM, with total height $H$, is partitioned along the $k_x$ direction into three functional sections: $l_1$ and $l_3$ serving as impedance-matching sections, flanking the central $l_2$ section, which is the main contributing region of the effective refractive index $n_{eff}$. The dielectric thickness gradually increases from 0 at the outer boundaries to a maximum thickness $h$ at the $l_2$ interfaces, maintaining a constant height h throughout the $l_2$ section. The effective refractive index $n_{eff}$ of the AIMM is controlled by adjusting the dielectric mixing ratio $m = 2h/H$. The quantitative relationship is expressed as:

$$n_{eff} = \frac{\int_0^{l_1} \sqrt{\frac{\varepsilon_0 \varepsilon_{rd}}{\varepsilon_0\left(\frac{m}{l_1}k_x\right) + \varepsilon_{rd}\left(1-\frac{m}{l_1}k_x\right)}}dk_x + \sqrt{\frac{\varepsilon_0 \varepsilon_{rd}}{\varepsilon_0 m + \varepsilon_{rd}(1-m)}} \cdot l_2 + \int_0^{l_3} \sqrt{\frac{\varepsilon_0 \varepsilon_{rd}}{\varepsilon_0\left(\frac{m}{l_3}k_x\right) + \varepsilon_{rd}\left(1-\frac{m}{l_3}k_x\right)}}dk_x}{L}, \quad (1)$$

where $\varepsilon_0$ represents the relative permittivity of air (approximated as 1 at microwave frequencies), and $L = l_1 + l_2 + l_3$ represents the total unit cell length along the $k_x$ -direction. As shown in Figs. 2(b) and (c), the effective refractive index $n_{eff}$ and transmittance $t$ of the AIMM are plotted across the 2-8 GHz under the parameters $l_1 = 0.02\lambda_0$, $l_2 = 0.02\lambda_0$, $l_3 = 0.02\lambda_0$, $H = 0.02\lambda_0$, and $\varepsilon_{rd} = 10$ (*e.g.*, $Al_2O_3$[29]), with the dielectric mixing ratio $m$ increasing from 0.25 to 0.75. The results reveal that the effective refractive index $n_{eff}$ of the AIMM varies continuously from 1.1 to 1.5, while the structure exhibits dispersionless characteristics over the 2-8 GHz bandwidth, with the transmittance $t$ remaining above 93%.

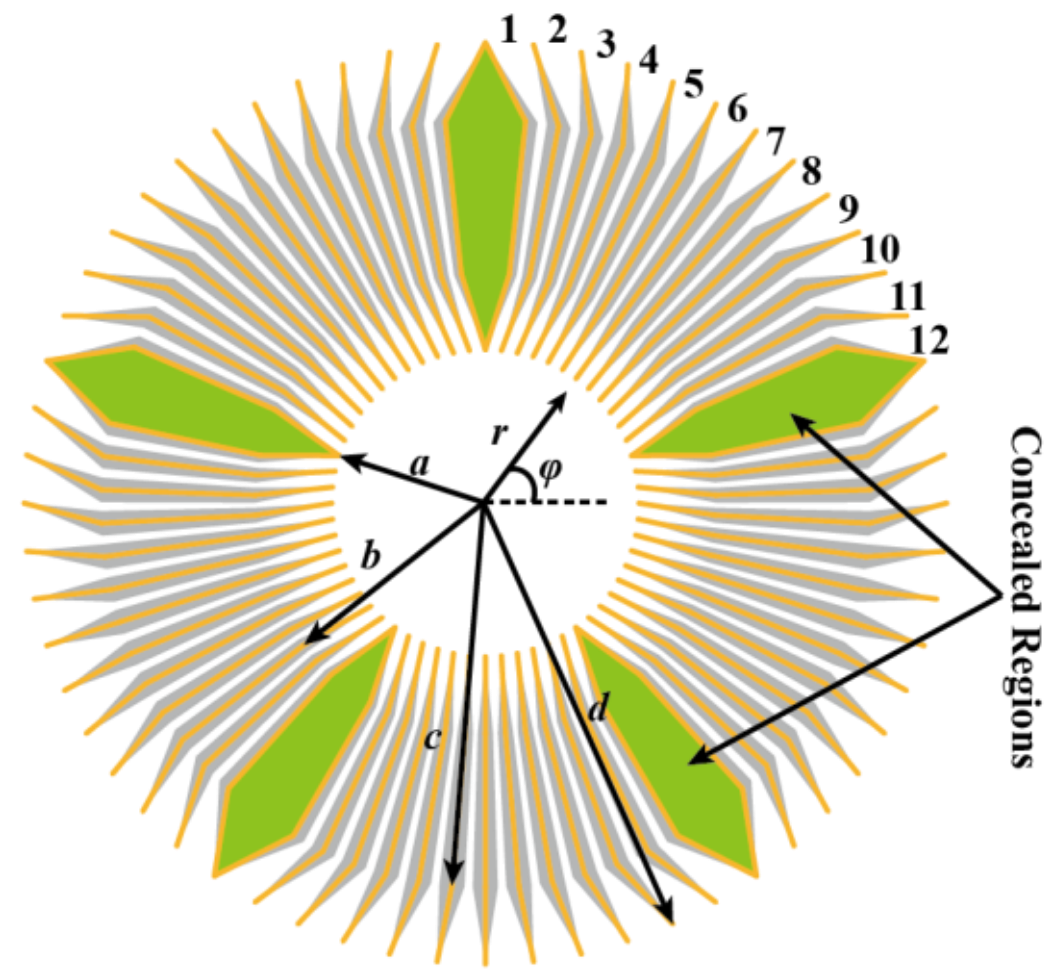

**Fig. 3.** The structural schematic diagram of SSC. The green areas represent the concealed regions.

Fig. 3 illustrates the SSC constructed from a series of AIMM within a cylindrical coordinate system. Electromagnetic waves emitted omnidirectionally from the SSC surrounded source are guided by the AIMM array to bypass obstacles while precisely reproducing their original wavefronts at the outside of SSC. Using fivefold rotational symmetry, we detail the design of a 72° sector (1/5 of the full structure), with identical remaining sectors generated via rotational duplication. First, transform the AIMM into cylindrical coordinates by converting $k_x$ to $r$ and $E_y$ to $\varphi$. As an example, set $a$ = 0.5$\lambda_0$, $b$ = 0.75$\lambda_0$, $c$ = 1.25$\lambda_0$, and $d$ = 1.5$\lambda_0$. The angle $H$ occupied by each channel linearly decreases from 6° at a to 5° at $b$, remains at 5° from $b$ to $c$, and then linearly increases back to 6° from $c$ to $d$. The selection of $H$ must satisfy the subwavelength condition at the highest operating frequency. Five hexagonal concealed regions surrounded by metal walls can be generated via rotational symmetry (the green areas in Fig. 3). Since the lengths of the channels are inconsistent, to ensure the wavefront shapes remains unchanged, the dielectric mixing ratio $m$ is adjusted to equalize the optical path length $di$ ($i$ = 1, 2, 3, ..., 12, representing the channel number) through the condition:

$$d_i = n_{eff}^i L_i = constant. \quad (2)$$

TABLE I. THE AIMM PARAMETERS FOR THE SSC IN FIG. 3

| Channel Number | $m$ | $l_1$ (mm) | $l_2$ (mm) | $l_3$ (mm) |
|---|---|---|---|---|
| 1 | 0.7054 | 15.411 | 30.000 | 16.957 |
| 2 | 0.7170 | 15.278 | 30.000 | 16.341 |
| 3 | 0.7262 | 15.170 | 30.000 | 15.831 |
| 4 | 0.7332 | 15.088 | 30.000 | 15.438 |
| 5 | 0.7379 | 15.034 | 30.000 | 15.170 |
| 6 | 0.7402 | 15.006 | 30.000 | 15.034 |
| 7 | 0.7402 | 15.006 | 30.000 | 15.034 |
| 8 | 0.7379 | 15.034 | 30.000 | 15.170 |
| 9 | 0.7332 | 15.088 | 30.000 | 15.438 |
| 10 | 0.7262 | 15.170 | 30.000 | 15.831 |
| 11 | 0.7170 | 15.278 | 30.000 | 16.341 |
| 12 | 0.7054 | 15.411 | 30.000 | 16.957 |

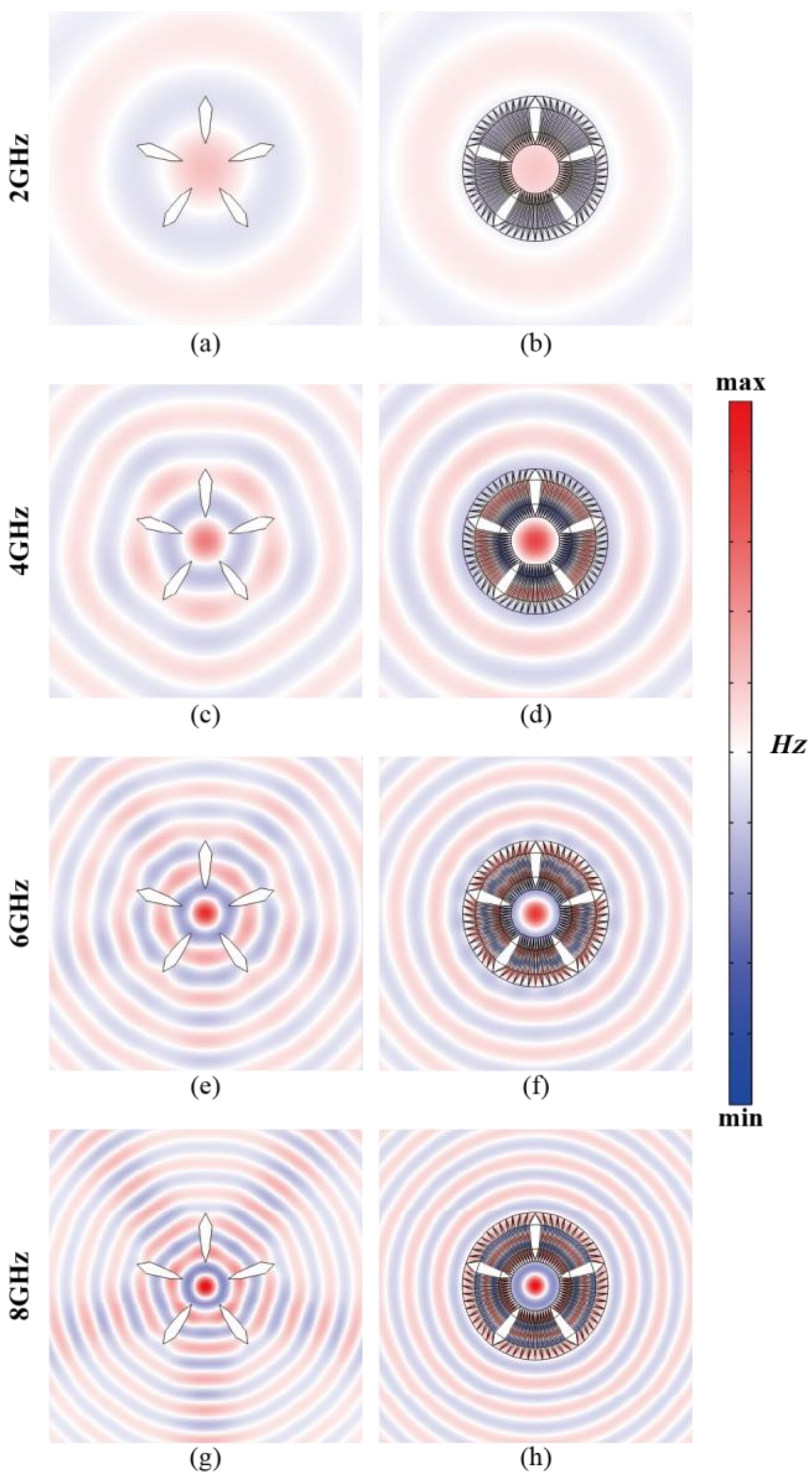

**Fig. 4.** Normalized magnetic field distribution at 2, 4, 6, 8 GHz from 2D numerical simulation with a centered magnetic current point source. (a), (c), (e), and (g) The reference. (b), (d), (f), and (h) The cloak surrounding the source.

The optical path length $d_i$ for all channels is set to 3/2$\lambda_0$. According to (1) and (2), the dielectric mixing ratio $m$ and structural parameters $l_1$, $l_2$, and $l_3$ of the AIMMs in each channel are calculated and summarized in Table I.

## III. RESULTS AND DISCUSSION

Fig. 4 presents numerical simulations of the SSC conducted with COMSOL Multiphysics (License #9406999) with its two-Dimensional (2D) Wave Optics module. The simulation domain is limited to a 6.5λ$_0$ × 6.5λ$_0$ square region surrounded by perfect matched layers of thickness 0.5λ$_0$, centered on a 1 $V$ magnetic point source. Figs. 4(a), (c), (e), and (g) show reference normalized magnetic field distributions with five PEC obstacles (matching the concealed regions geometry) surrounding the source at 2, 4, 6, 8 GHz. Strong radiation pattern distortions caused by PEC scattering are observed, exacerbating at shorter wavelengths. Figs. 4(b), (d), (f), and (h) demonstrate the source surrounded by SSC at 2, 4, 6, 8 GHz. Here, electromagnetic waves emitted omnidirectionally from the source are guided around the obstacles and reproduce cylindrical wavefronts at the outer boundary of SSC across the entire 6 GHz bandwidth. These results validate that the SSC effectively protects the radiation pattern of broadband sources from scattering by obstacles, regardless of emission direction or operating frequency within the designed bandwidth.

To further validate the applicability of the SSC, we conduct 2D numerical simulations to verify its effectiveness when the point source is offset from the center of the SSC. As shown in Fig. 5, the magnetic point source is offset to the right by 0.25$\lambda_0$ from the center. The reference normalized magnetic field distributions with five PEC obstacles (matching the concealed regions geometry) surrounding the source show significant radiation pattern distortions caused by scattering from the PEC obstacles (see Fig. 5(a), (c), (e), and (g)). However, when the source is surrounded by the SSC, the electromagnetic waves emitted by the source are guided around the concealed regions and reproduce cylindrical wavefronts at the outer boundary of SSC across the entire operating bandwidth, protecting the radiation pattern of broadband sources from scattering by obstacles (see Fig. 5(b), (d), (f), and (h)). Notably, the distribution of electromagnetic waves outside the SSC is also offset to the right by $0.25\lambda_0$ relative to the case where the source is at the center of the SSC. This demonstrates that the SSC remains effective for electromagnetic wave sources located anywhere within its inner boundary.

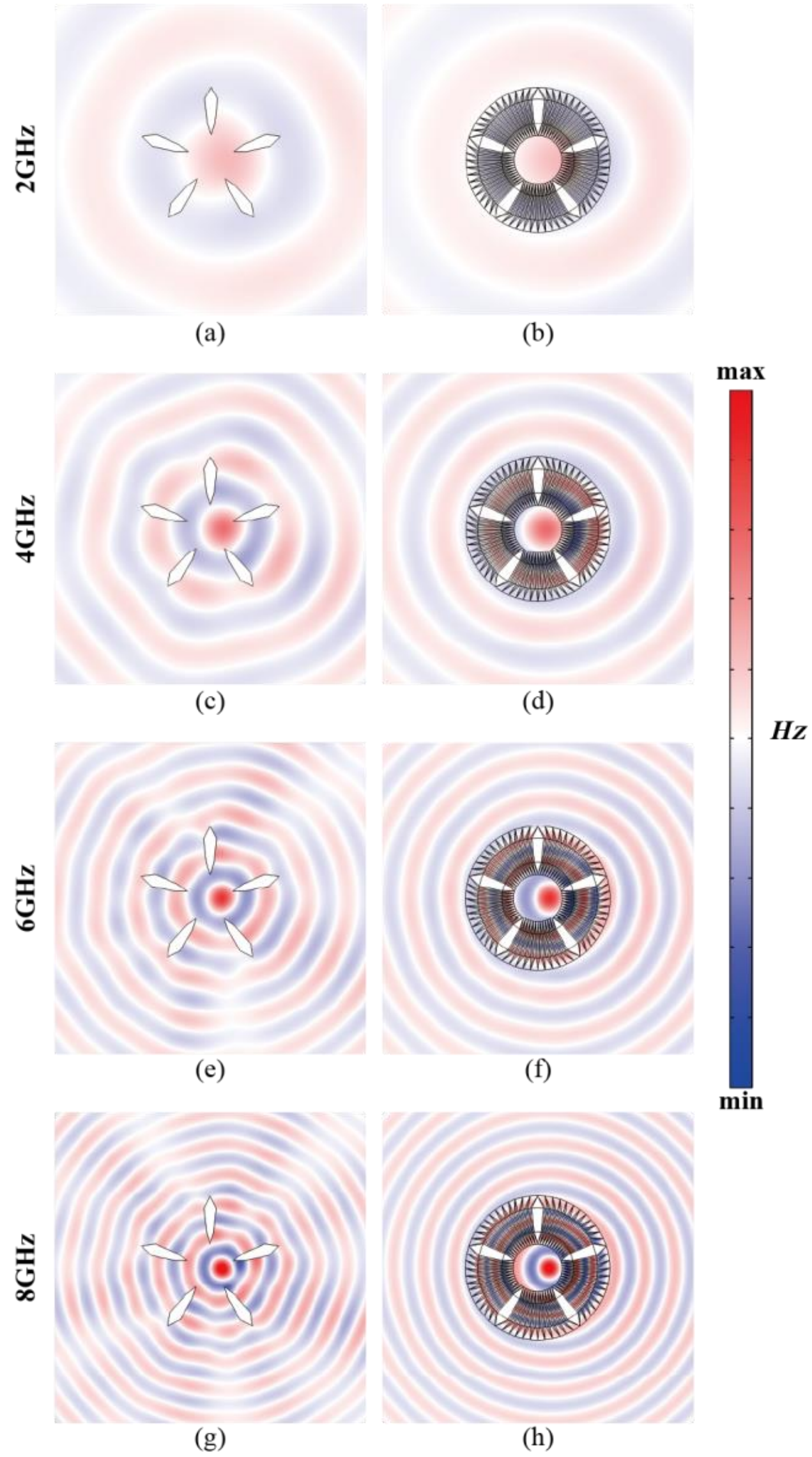

**Fig. 5.** Normalized magnetic field distribution at 2, 4, 6, 8 GHz from 2D numerical simulation with the magnetic current point source positioned 15 mm off-center. (a), (c), (e), and (g) The reference. (b), (d), (f), and (h) The cloak surrounding the source.

## IV. Conclusion

In conclusion, in this work, we propose a dispersionless air-impedance-matched metamaterial over the 2-8 GHz bandwidth that achieves an adjustable effective refractive index ranging from 1.1 to 1.5, with transmittance maintained above 93%. Based on the AIMM, we design an SSC which surrounds the electromagnetic wave source to create five concealed regions. TM electromagnetic waves emitted by the source in any propagation direction can be guided by the SSC to bypass the concealed regions and reproduce the original wavefronts outside the SSC at 2-8 GHz. Thus, it protects the radiation pattern from being distorted by scattering from obstacles. We further validate the applicability of its functionality. Even when the electromagnetic wave source is not strictly centered within the SSC but offset by a certain distance, the SSC still protects the radiation patterns from scattering by obstacles, and the reproduced electromagnetic waves are similarly offset by that distance. Our work demonstrates significant potential for enhancing the integration density of integrated circuits and improving the operational stability of communication systems.